\documentstyle[aps,eqsecnum,epsf]{revtex}
\begin{document}
\draft

\wideabs{
\title{
The Transition from Inspiral to Plunge for a Compact Body in a 
Circular Equatorial Orbit Around a Massive, Spinning Black Hole 
}
\author{Amos Ori,$^{(1)}$ and Kip S.  Thorne$^{(2)}$
}
\address{
$^{(1)}$Department of Physics, Technion---Israel Institute of Technology,
Haifa, 32000, Israel \\
$^{(2)}$Theoretical Astrophysics, California Institute of Technology, Pasadena,
CA 91125 \\
}
\date{Received 21 February 2000}
\maketitle
\begin{abstract}

There are three regimes of gravitational-radiation-reaction-induced
inspiral for a
compact body with mass $\mu$, in a circular, equatorial orbit around a  
Kerr black hole with mass $M\gg\mu$:  
(i) The {\it adiabatic inspiral regime}, in
which the body gradually descends through a sequence of circular, geodesic
orbits. (ii) A {\it transition regime}, near the innermost 
stable circular orbit (isco). (iii) The {\it plunge regime}, in which 
the body travels on a geodesic from slightly below the isco into the 
hole's horizon.  This paper gives an analytic treatment of the transition
regime and shows that, 
with some luck, gravitational waves from the transition might
be measurable by the space-based LISA mission.

\end{abstract}
}
\narrowtext

\section{INTRODUCTION AND SUMMARY}
\label{sec:Intro}

The space-based Laser Interferometer Space Antenna (LISA) \cite{lisa}, 
if it flies,
is likely to detect and study the gravitational waves from white dwarfs,
neutron stars and small black holes with masses $\mu \agt 1 M_\odot$,
spiraling into Massive 
($M\sim 10^5$--$10^8 M_\odot \gg \mu$) 
black holes in the nuclei of distant galaxies
\cite{sigurdssonrees,sigurdsson,finnthorne}.  In preparation for these
studies, it is necessary to understand, theoretically, the 
radiation-reaction-induced evolution of the inspiral orbits, and the 
gravitational waveforms that they emit.

Regardless of an orbit's shape and orientation, 
when $\mu \ll M$ 
the orbital evolution can be divided into three regimes: 
(i) The {\it adiabatic inspiral
regime}, in which the body gradually descends through a sequence of geodesic
orbits with gradually changing ``constants'' of the motion $E = ($energy), $L
= ($polar component of angular momentum), and $Q =($Carter constant). 
(ii) A {\it transition
regime}, in which the character of the orbit gradually changes from
inspiral to plunge.  (iii) A {\it plunge
regime}, in which the body plunges into the horizon along a geodesic with
(nearly) unchanging $E$, $L$ and $Q$.

The plunge regime, being (essentially) ordinary geodesic motion, is well 
understood; and 
the adiabatic inspiral regime is the focus of extensive current research (see,
e.g.\ \cite{finnthorne,hughes,barackori}).  By contrast, so far as we are aware,
there have been no publications dealing with the transition regime.

In 1990-91, we carried out an initial exploration of the transition
regime for the special case of circular, equatorial orbits; but due to 
the press of other projects we did not publish it.
Now, with prospects for LISA looking good, we have resurrected our work
and present it here, in the context of LISA.  

We begin, in Sec.\ \ref{sec:InspiralPlunge}, by summarizing some key,
well-known details of the inspiral and plunge regimes.  Then in Sec.\ 
\ref{subsec:Qualitative} we present a qualitative picture of the transition
from inspiral to plunge, based on the motion of a particle in a slowly changing
effective potential (Fig.\  \ref{figeffpotl}).  With the aid of this
qualitative picture, in Sec.\ \ref{subsec:EOM} 
we derive a non-geodesic equation of motion for 
the transition regime, and in Sec.\ \ref{subsec:Solution} we
construct the solution
to that equation of motion (Figs.\ 
\ref{figtransition1} and \ref{figtransition2}).  Then in Sec.\ 
\ref{sec:GW}, with the aid of our
solution, we estimate the gravitational-wave signal strength from the
transition regime 
and the signal-to-noise ratio that it would produce in LISA.  We conclude
that, with some luck, LISA may be able to detect and study the transition
waves.
In Sec.\ \ref{sec:Conclusions} we make concluding remarks
about the need for further research.

\section{ADIABATIC INSPIRAL AND PLUNGE}
\label{sec:InspiralPlunge}

Throughout this paper we use Boyer-Lindquist coordinates $(t,r,\theta,\phi)$ 
\cite{mtw} for
the massive hole's Kerr metric, and we use geometrized units, with $G=c=1$.  
The hole's mass is $M$ and the inspiraling body's mass is 
$\mu \equiv \eta M$.  
We use $M$ and $\mu$ to construct dimensionless versions
(denoted by tildes) of many
dimensionfull quantities; for example, $\tilde r = r/M$, and $\tilde t = t/M$. 
The hole's dimensionless spin parameter is 
$a \equiv ($spin angular momentum$)/M^2$ (with $-1<a<+1$). 
The body moves around its 
circular, equatorial
orbit in the $+\phi$ direction, so $a>0$
corresponds to an orbit that is prograde relative to the hole's spin,
and $a<0$ to a retrograde orbit.

When the inspiraling body is not too close to the innermost stable circular
orbit (isco), it moves on a 
circular geodesic orbit with dimensionless angular velocity \cite{bpt}
\begin{equation}
\tilde\Omega \equiv 
M\Omega = {d\phi\over d\tilde t} = {1\over{\tilde r^{3/2}+a}}
\label{Omega}
\end{equation}
(where $\phi$ is angle around the orbit) and with orbital energy \cite{bpt} 
\begin{equation}
E = - \eta M {1-2/\tilde r + a/\tilde r^{3/2} \over \sqrt{1-3/\tilde r +
2a/\tilde r^{3/2}}}\;.
\label{Er}
\end{equation}

As it moves, the body radiates
energy into gravitational waves at a rate given by \cite{finnthorne}
\begin{equation}
\dot E_{\rm GW}  = - \dot E
= {32\over 5} \eta^2\tilde\Omega^{10/3}\dot {\cal E},
\label{dotE}
\end{equation}
where $\dot {\cal E}$ is a general relativistic correction to the Newtonian,
quadrupole-moment formula (Table II of Ref.\ \cite{finnthorne}).  This energy
loss causes the orbit to shrink adiabatically at a rate given by 
\begin{equation}
{dr\over dt} = {- \dot E_{\rm GW}\over dE/dr}\;.
\label{drdt}
\end{equation}

The inspiral continues adiabatically until the body nears the isco, which
is at the dimensionless radius 
$\tilde r_{\rm isco} = r_{\rm isco}/M$ given by \cite{bpt}
\begin{eqnarray}
\tilde r_{\rm isco} = &&3+Z_2-{\rm sign}(a)
[(3-Z_1)(3+Z_1+2Z_2)]^{1/2}\;,\nonumber\\
&&Z_1\equiv 1+(1-a^2)^{1/3}[(1+a)^{1/3} + (1-a)^{1/3}]\;,\nonumber\\
&&Z_2\equiv (3a^2 + {Z_1}^2 )^{1/2}\;; \label{rms}
\end{eqnarray}
cf.\ Table \ref{Table.tbl}.
The circular geodesic orbit at the isco has dimensionless
angular velocity (Table \ref{Table.tbl}), energy, and
angular momentum given by \cite{bpt,mtw}
\begin{equation}
\tilde\Omega_{\rm isco} \equiv M\Omega
= {1\over {\tilde r_{\rm isco}}^{3/2} +a}\;,
\label{Omegams}
\end{equation}
\begin{equation}
\tilde E_{\rm isco} \equiv {E_{\rm isco}\over\mu} = {E_{\rm isco}\over\eta M}
= {1-2/\tilde r_{\rm isco} + a/{\tilde r_{\rm
isco}}^{3/2}\over
\sqrt{1-3/\tilde r_{\rm isco} + 2a/{\tilde r_{\rm isco}}^{3/2} } }\;,
\label{Ems}
\end{equation}
\begin{equation}
\tilde L_{\rm isco} \equiv {L_{\rm isco}\over\mu M} = 
{L_{\rm isco}\over\eta M^2}
= {2\over\sqrt{3\tilde r_{\rm isco}}}\left(3\sqrt{\tilde
r_{\rm isco}} -2a\right)\;.
\label{Lms}
\end{equation}

As the body nears the isco, its inspiral gradually ceases to be adiabatic 
and it enters the transition regime (Sec.\ \ref{sec:Transition}).  
Radiation reaction (as controlled by $\dot E_{\rm GW}$)
continues to drive the orbital evolution throughout the transition regime, 
but gradually becomes unimportant
as the transition ends and pure plunge takes over.  

The plunge is described to high accuracy by reaction-free geodesic motion;
Eqs.\ (33.32) of Ref.\ \cite{mtw}. 
Up to fractional corrections
of order $\eta^{4/5}$, the orbital energy and angular
momentum of the plunging body are equal to $E_{\rm isco}$ and $L_{\rm isco}$
throughout the plunge.
[cf.\ Eq.\ (\ref{final-values}) below].

\section{THE TRANSITION FROM ADIABATIC INSPIRAL TO PLUNGE}
\label{sec:Transition}

\subsection{Qualitative Explanation of Transition}
\label{subsec:Qualitative}

As the body nears its innermost stable circular orbit, $r=r_{\rm isco}$,
the adiabatic approximation begins to break down.  This breakdown can
be understood in terms of the effective potential, which governs 
{\it geodesic} radial motion via the equation
\begin{equation}
\left({d\tilde r\over d\tilde \tau}\right)^2 =  
\left({d r\over d \tau}\right)^2 = 
\tilde E^2 -V(\tilde r, \tilde E, \tilde L)\;,
\label{rgeo}
\end{equation}
where $\tilde E\equiv E/\mu = E/(\eta M)$, $\tilde L\equiv L/(\mu M)
=L/(\eta M^2)$, and $\tilde\tau
\equiv \tau/M$ 
are the body's dimensionless energy, angular momentum, and proper time.
The explicit form of the effective potential can be
inferred from Eqs.\ (33.32) and (33.33) of MTW\cite{mtw}:  
\begin{eqnarray}
V(\tilde r, \tilde E, \tilde L) &=& \tilde E^2 - 
{1\over \tilde r^4}\left([\tilde
E(\tilde r^2+a^2)-\tilde La]^2\right.\nonumber\\
&&- \left.(\tilde r^2-2\tilde r +a^2)[\tilde r^2+(\tilde L-\tilde E a)^2]
\right)\;.
\label{VrEL}
\end{eqnarray}
For a Schwarzschild black hole, this reduces to
\begin{equation}
V(\tilde r, \tilde E, \tilde L) =
\left(1-{2\over\tilde r}\right)\left(1+{\tilde L^2\over \tilde
r^2}\right)\quad\hbox{for }a=0
\end{equation}
(cf.\ Eq. (25.16) of MTW \cite{mtw}).

\begin{figure}
\epsfxsize=3.2in\epsfbox{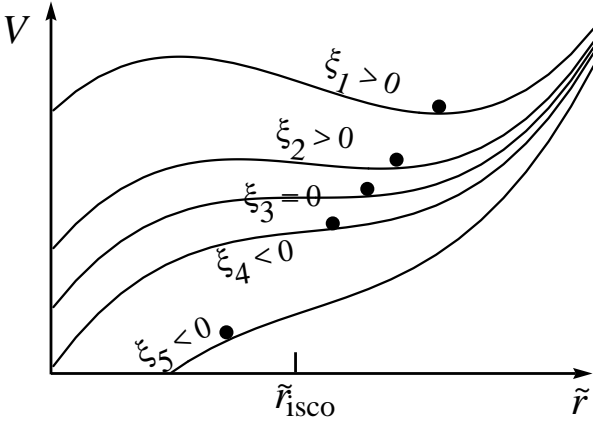}
\caption{The gradually changing
effective potential $V(\tilde r,\xi)$ for radial geodesic motion.  Each curve
is for a particular value of $\xi \equiv \tilde L - \tilde L_{\rm
isco}$.  As $\xi$ decreases due to radiation reaction, the body, depicted
by the large dot, at first remains at the minimum of the effective potential
($\xi_1$; ``adiabatic regime''). However, as $\xi$ nears zero (at
$\xi\simeq\xi_2$), the body
cannot keep up with the rapid inward motion of the minimum; it lags
behind in a manner described by the transition-regime analysis of Sec.\
\protect\ref{sec:Transition}.  At $\xi\simeq\xi_5$ the effective potential has
become so steep that radiation reaction is no longer important, the
transition regime ends, and the body plunges toward the black hole with
nearly constant energy and angular momentum.
\label{figeffpotl}
}
\end{figure}

Throughout the inspiral and transition regimes, the body moves along a
nearly circular orbit; its change of radius during each circuit around 
the hole is $\Delta
r \ll r$.  (Only after the body is well into its final plunge toward the hole
does $\Delta r$ become comparable to $r$.)  This near-circular motion
guarantees that the ratio of the energy radiated to angular momentum 
radiated is equal to the body's orbital angular velocity 
\cite{zeldovichnovikov}:
\begin{equation}
{d\tilde E\over d\tilde\tau} = \tilde\Omega{d\tilde L\over d\tilde\tau}\;.
\label{universal}
\end{equation}
Correspondingly, in and near the transition regime, which occupies
a narrow range of radii around 
$\tilde r_{\rm isco}$, 
the body's energy and angular 
momentum
are related by\footnote{
In reality,
finite-mass-ratio effects, including those discussed in the paragraph
preceding Eq.\ (\ref{defR}) below, will alter these energy-angular-momentum
relations by amounts that scale as the first and higher powers of $\eta$.
For example, in going from Eq.\ (\ref{universal}) to (\ref{defxi}), there can
be an integration constant $\delta \tilde E$ (which scales as $\eta$ or some
higher power) so $\tilde E = \tilde E_{\rm isco} + \delta \tilde E +
\tilde \Omega_{\rm isco} \xi$. 
In the presence of such effects, we {\it redefine} $\tilde r_{\rm isco}$, 
$\tilde 
E_{\rm isco}$, and $\tilde L_{\rm isco}$ to be the values of these parameters
at which the $\eta$-corrected
$V(\tilde r, \tilde E, \tilde L)$ has a flat inflection point, as in
Fig.\ \ref{figeffpotl},
and $\tilde \Omega_{\rm isco}$ to be the orbital angular velocity at this
$\tilde r_{\rm isco}$.
Then Eqs.\ (\ref{defxi}) remain valid even for finite
mass ratio $\eta$. 
}
\begin{equation}
\tilde E = \tilde E_{\rm isco} + \tilde \Omega_{\rm isco} \xi\;,\quad 
\tilde L = \tilde L_{\rm isco} + \xi\;.
\label{defxi}
\end{equation}
By combining Eqs.\ (\ref{defxi}) and (\ref{VrEL}), we can regard the
body's effective potential as a function of $\tilde r$ and the
difference $\xi \equiv \tilde L - \tilde L_{\rm isco}$ of its orbital
angular momentum from that of the isco.  

Figure \ref{figeffpotl} shows $V(\tilde r, \xi)$ for a sequence of angular
momenta $\xi_1,\ldots,\xi_5$ around 
$\xi =0$.  As $\xi$ decreases to $\xi = 0$, 
the minimum of the potential flattens
out and disappears; and just when it is disappearing, the 
minimum's radius $r_{\rm min}$
is moving inward at an infinite rate: $dr_{\rm min}/d\xi
\rightarrow\infty$ as $\xi\rightarrow 0$.

In the adiabatic regime of large $\xi$, the body sits always at the
minimum of the effective potential.  Its orbit is a slowly shrinking
circle, guided inward by the motion of the minimum.  As $\xi$
nears zero and the minimum's inward speed grows large, the body's 
inertia prevents
it from continuing to follow the minimum.  The body begins to
lag behind, as depicted at $\xi = \xi_2$
in Fig.\ \ref{figeffpotl}.  This lag invalidates the adiabatic inspiral
analysis of Sec.\ \ref{sec:InspiralPlunge} and initiates the transition
regime.  

As $\xi$ continues to decrease, there comes a point (near
$\xi_5$ in Fig.\ \ref{figeffpotl}) at which the effective potential 
has become so steep that its
inward force on the body dominates strongly over radiation reaction.
There the transition regime ends, and the body begins to plunge inward 
rapidly on 
a nearly geodesic orbit with nearly constant $\tilde E$ and $\tilde L$. 
The objectives of the following subsections are to derive a set of 
equations describing
the transition regime (Sec.\ \ref{subsec:EOM}), and 
show how the transition matches smoothly onto the
adiabatic regime at large positive $\xi$ and to the plunge regime at large
negative $\xi$ (Sec.\ \ref{subsec:Solution}).

\subsection{Equation of Motion for Transition Regime}
\label{subsec:EOM}

Throughout the transition regime, because 
the body moves on a nearly circular orbit with radius close to
$r_{\rm isco}$, and because the body's small mass $\mu \equiv \eta M \ll M$
keeps its radiation reaction weak, its angular velocity remains very close to
$\Omega_{\rm isco}$ 
\begin{equation}
{d\phi\over d\tilde t} \equiv \tilde\Omega \simeq 
{\tilde\Omega_{\rm isco}}\;,
\label{dphidt}
\end{equation}
and its proper time ticks at very nearly the standard 
isco rate
\begin{equation}
{d\tilde\tau\over d\tilde t} \simeq 
\left({d\tilde\tau\over d\tilde t}\right)_{\rm isco} =
{ \sqrt{1-3/\tilde r_{\rm isco} + 2a/{\tilde r_{\rm
isco}}^{3/2}}\over 1+a/{{\tilde r_{\rm isco}}}^{3/2}}\;;
\label{dtaudt}
\end{equation}
cf.\ Eq.\ (5.4.5a) of \cite{novikovthorne}. 
 
This nearly circular motion at $\tilde r \simeq \tilde r_{\rm isco}$ 
produces gravitational waves which carry off angular momentum and energy
at very nearly the
same rate as they would for circular geodesic motion at $\tilde r_{\rm
isco}$.  This means that $\tilde E$ and $\tilde L$ evolve in accord with
Eqs.\ (\ref{defxi}), where
\begin{equation}
{d\xi\over d\tilde \tau} = -\kappa\eta\;, 
\label{dxidtau}
\end{equation}
and
\begin{equation}
\kappa = {32\over5}{\tilde\Omega_{\rm isco}}^{7/3}
{1+a/{\tilde r_{\rm
isco}}^{3/2}\over\sqrt{1-3/\tilde r_{\rm isco} + 2a/{\tilde r_{\rm
isco}}^{3/2}}}\dot {\cal E}_{\rm isco}\;;
\label{kappa}
\end{equation}
cf.\ Eqs.\ (\ref{dotE}), (\ref{defxi}), (\ref{dtaudt}), and Table
\ref{Table.tbl}.  It is the
smallness of $\eta \equiv \mu/M$ (e.g., $\eta = 10^{-5}$ for the
realistic case of a $10 M_\odot$ black hole spiraling into the 
$10^6M_\odot$ black hole) 
that makes the angular momentum $\xi$ evolve very slowly
and keeps the body in a nearly circular orbit throughout the transition
regime
[cf.\ the factors of $\eta$ that appear in 
Eqs.\ (\ref{dxidtau}) and (\ref{XTDef})---which with 
Eqs.\ (\ref{defR}) and (\ref{eomdimensionless}) 
imply $d\tilde r/d\tilde \tau \propto \eta^{3/5}$.]

In the transition regime, the body's radial motion is 
described by the geodesic equation of motion with a radial 
self force 
per unit mass\footnote{This radial self force, like the radiation
reaction force that drives the inspiral, is produced by interaction of
the body with its own gravitational field---that field having been influenced
by the black hole's spacetime geometry; see, e.g., Ref.\ \cite{barackori}.  
The contravariant radial component of the self force, with
dimensionality restored using $r = M \tilde r$ and $\tau = M \tilde \tau$,
is 
$(dp^r/d\tau)_{\rm self} 
= (\mu d^2 r/d\tau^2)_{\rm self} = 
(\mu/M) (d^2\tilde r /d\tilde\tau^2)_{\rm
self} 
= \eta^2 \tilde F_{\rm self}$.
} 
$\eta \tilde F_{\rm self}$ 
inserted on the right-hand side:
\begin{equation}
{d^2 \tilde r\over d\tilde\tau^2} = -{1\over2}{\partial V(\tilde r,
\xi)\over\partial \tilde r} + \eta\tilde F_{\rm self}\;.
\label{eom}
\end{equation}
(We write it as $\eta \tilde F_{\rm self}$ because its magnitude
is proportional to $\eta = \mu/M$.)

The radial 
self force
$\eta\tilde F_{\rm self}$ is
nondissipative (since it has hardly any radial velocity with which to couple). 
This contrasts with
the $\phi$-directed
radiation-reaction force, which couples to the orbital angular
velocity to produce a shrinkage of
the body's angular momentum [Eq.\ (\ref{dxidtau})] and a corresponding
decrease of its energy, $d\tilde E/d\tilde\tau =
\tilde\Omega d\xi/d\tilde\tau$.  Because the radial force is nondissipative,
it is of little importance.  It can be absorbed into the nondissipative
effective potential term $-{1\over2}\partial V/\partial\tilde r$ in the 
equation of motion.  Doing so will not change the general character of
the effective potential, as depicted in Fig.\ \ref{figeffpotl}; it will
merely change, by fractional amounts proportional to $\eta$, the 
various parameters that characterize the effective potential: 
the location $\tilde r_{\rm isco}$ of the innermost stable circular orbit 
(at which the $\xi=0$ effective potential curve has its inflection point),
the values at the isco of the orbital energy and angular momentum 
$\tilde E_{\rm isco}$ and $\tilde L_{\rm isco}$, and the constant $\alpha$
defined below.  
There will be other ${\rm O}(\eta)$ changes 
in $\tilde r_{\rm isco}$, $\tilde E_{\rm isco}$, $\tilde L_{\rm isco}$
and $\alpha$
caused by the body's own perturbation of
the hole's spacetime geometry\cite{kww,papproximates,buonannodamour}.  
In this paper, we shall ignore
all such changes, and correspondingly we shall neglect the radial
self force
$\eta \tilde F_{\rm self}$.  

We shall describe the body's location in the transition regime by 
\begin{equation}
R\equiv \tilde r-\tilde r_{\rm isco}\;.
\label{defR}
\end{equation}
Throughout the 
transition regime both $R$ and $\xi$ are small, and correspondingly the
effective potential can be expanded in powers of $R$ and $\xi$.  Up
through cubic terms in $R$ and linear terms in $\xi$ (the order needed
for our analysis), the effective potential takes the form
\begin{equation}
V(R,\xi) = {2\alpha\over3}R^3-2\beta R\xi +\hbox{constant}\;,
\label{VRxi}
\end{equation}
where $\alpha$ and $\beta$ are positive constants that we shall evaluate
below.
Note that for $\xi=0$, this is a simple cubic potential with inflection 
point at $R=0$,
i.e.\ at $\tilde r=\tilde r_{\rm isco}$; 
and note that for $\xi >0$, it acquires 
a maximum and a minimum, while for $\xi <0$ it is monotonic; cf.\ Fig.\
\ref{figeffpotl}.  By inserting Eq.\ (\ref{VRxi}) into Eq.\ (\ref{eom}),
setting $\tilde r = \tilde r_{\rm isco} + R$, 
and neglecting the radial 
self force
or absorbing it into
$\tilde r_{\rm isco}$, $\alpha$ and $\beta$
as described above,
we obtain the
following radial equation of motion: 
\begin{equation}
{d^2 R\over d\tilde\tau^2} = -\alpha R^2+\beta\xi\;.
\label{eom1}
\end{equation}
By then setting $\tilde\tau \equiv 0$ at the moment when $\xi=0$ and using
Eq.\ (\ref{dxidtau}) for the rate of change of $\xi$, 
so
\begin{equation}
\xi = - \eta \kappa \tilde \tau\;,
\label{xitau}
\end{equation}
we bring our
equation of motion into the form
\begin{equation}
{d^2 R\over d\tilde\tau^2} = -\alpha R^2-\eta\beta\kappa\tilde\tau\;.
\label{eom2}
\end{equation}
We shall explore the consequences
of this equation of motion in the next subsection, but first we shall
deduce the values of $\alpha$ and $\beta$. 

The constants $\alpha$ and $\beta$ can be evaluated from the following
relations, which follow directly from (\ref{VRxi}), 
(\ref{defR}) 
and (\ref{defxi}):
\begin{equation}
\alpha = {1\over4}
\left({\partial^3 V(\tilde r, \tilde E, \tilde L)\over\partial 
\tilde r^3}\right)_{\rm isco}\;,
\label{alpha}
\end{equation}
\begin{equation}
\beta = - {1\over2}
\left( {\partial^2 V(\tilde r, \tilde E, \tilde L)\over 
\partial \tilde L \partial \tilde r} + \tilde\Omega 
{\partial^2 V(\tilde r, \tilde E, \tilde L)\over 
\partial \tilde E \partial \tilde r} \right)_{\rm isco}\;.
\label{beta}
\end{equation}
By inserting expression (\ref{VrEL}) into these relations, we obtain 
$\alpha$ and $\beta$ in the limit $\eta\equiv\mu/M \rightarrow 0$:
\begin{eqnarray}
\alpha &=& {3\over\tilde r_{\rm isco}^6}
\left(\tilde r^2 + 2[a^2(\tilde E^2-1) - \tilde
L^2]\tilde r + 10(\tilde L - a\tilde E)^2\right)_{\rm isco}\nonumber\\ 
&=&{1\over1296}\quad\hbox{for }a=0\;,
\label{alpha1}\\
\beta &=& {2\over \tilde r_{\rm isco}^4}
\left( (\tilde L - a^2\tilde E\tilde\Omega)\tilde r - 3(\tilde L - a
\tilde E)(1-a\tilde\Omega)\right)_{\rm isco}\nonumber\\
&=&{1\over 36\sqrt3}\quad\hbox{for }a=0\;.
\label{beta1}
\end{eqnarray}
Here $\tilde r_{\rm isco}$ and $\tilde\Omega_{\rm isco}$ are
given by Eqs.\ (\ref{rms}) and (\ref{Omegams}); and $\tilde
L_{\rm isco}$ and $\tilde E_{\rm isco}$ are expressed in terms of $\tilde r_{\rm
isco}$ by Eqs.\ (\ref{Ems}) and (\ref{Lms}).
Numerical values of $\alpha$ and $\beta$, computed from these equations,
are tabulated in Table \ref{Table.tbl}.

\subsection{Solution for Motion in the Transition Regime}
\label{subsec:Solution}

The equation of motion in the transition regime, Eq.\  (\ref{eom2}), can be
converted into dimensionless form by setting
\begin{equation}
R = \eta^{2/5} R_o X\;, \quad \tilde\tau = \eta^{-1/5} \tau_o T\;,
\label{XTDef}
\end{equation}
where
\begin{equation}
R_o = (\beta\kappa)^{2/5}\alpha^{-3/5}\;, \quad 
\tau_o = (\alpha\beta\kappa)^{-1/5}\;;
\label{RotauoDef}
\end{equation}
cf.\ Table \ref{Table.tbl}.
The resulting dimensionless equation of motion is
\begin{equation}
{d^2 X \over dT^2} = - X^2 - T\;.
\label{eomdimensionless}
\end{equation}

We seek the unique solution of this differential equation which, at
early times $T\ll -1$, joins smoothly onto the adiabatic inspiral 
solution of Sec.\ \ref{sec:InspiralPlunge}.  In that adiabatic inspiral,
the orbit is the circle at the minimum of the effective potential of
Fig.\ \ref{figeffpotl} and Eq.\ (\ref{VRxi}), i.e., the circle at $R =
\sqrt{\beta\xi/\alpha} = \sqrt{-\beta\kappa\eta\tau/\alpha}$, which
translates into
\begin{equation}
X = \sqrt{-T} \quad \hbox{for adiabatic inspiral near the isco.}  
\label{XTAdiabatic}
\end{equation}

\begin{figure}
\epsfxsize=3.2in\epsfbox{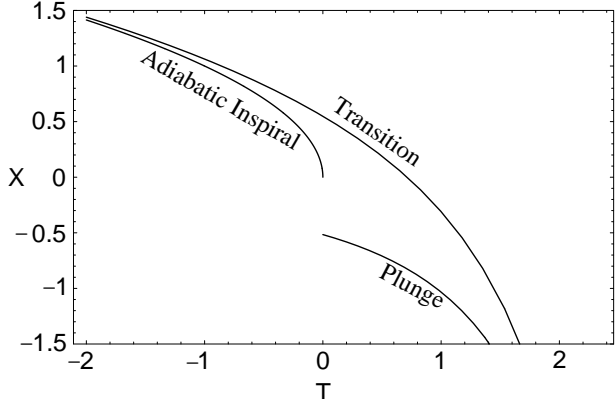}
\caption{
Dimensionless orbital radius $X$ as a function of dimensionless proper
time $T$ for an orbit near the isco.  {\it Adiabatic Inspiral:} The analytic
solution (\protect\ref{XTAdiabatic}) 
for adiabatic inspiral outside the 
isco. {\it Transition:} The numerical solution to the dimensionless equation 
of motion (\protect\ref{eomdimensionless}) 
for the transition regime in the vicinity
of the isco.  {\it Plunge:} The analytic solution 
(\protect\ref{XTPlunge}) for the orbital plunge inside the isco. 
}
\label{figtransition1}
\end{figure}

\begin{figure}
\epsfxsize=3.2in\epsfbox{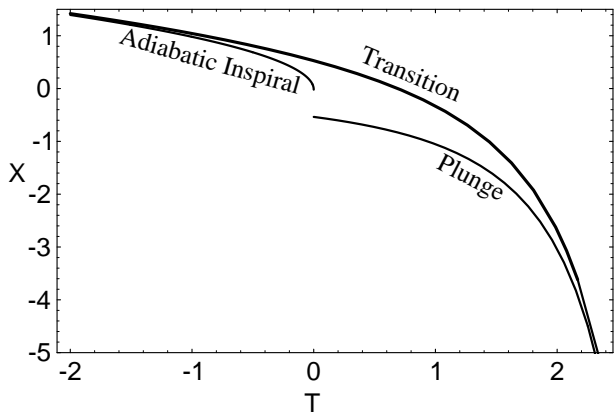}
\caption{Same as Fig.\ \protect\ref{figtransition1}, but drawn on a different
scale.
}
\label{figtransition2}
\end{figure}

We have not been able to find an analytic formula for the solution to
the equation of motion (\ref{eomdimensionless}) that joins smoothly onto this
adiabatic solution, but it is easy to construct the unique solution
numerically.  It is plotted in Figures \ref{figtransition1} and
\ref{figtransition2}, along with the adiabatic inspiral solution 
(\ref{XTAdiabatic}) and the plunge solution [Eq.\ (\ref{XTPlunge}) below].

The transition solution is well approximated by adiabatic inspiral
at times $T < -1$, but at $T>-1$ it deviates from adiabatic inspiral and
evolves smoothly into a plunge. 
The solution diverges ($X\rightarrow - \infty$) at a finite time
$T=T_{\rm plunge} \simeq 3.412$.\footnote 
{The divergence of $X$ at $T=T_{\rm plunge}$ does not imply 
a divergence of $r$ or any other physical quantity. 
Rather, it marks the breakdown of the transition approximation at 
very large values of $|X|$, $|X|\agt X_{\rm break} \sim \eta^{-2/5}$;
[cf.\ Eq.\ (\ref{XTDef})].
More specifically, when $-X \agt X_{\rm break}$, 
higher-order terms in the Taylor expansion (\ref{VRxi}) 
become important and stop the divergence.  Well before this
(in fact, throughout the range 
$1\ll-X\ll X_{\rm break}$) both the transition approximation and the free-fall
approximation ($\tilde E = \tilde E_{\rm final} = \hbox{constant}$, 
$\tilde L = \tilde L_{\rm final} = \hbox{constant}$) are valid, so these
two approximations can be matched in this regime to obtain a solution valid
all the way down to the horizon. 
The same type of breakdown also occurs at the other asymptotic limit
$+X\agt  X_{\rm break}$: The transition regime's 
adiabatic-inspiral equation (\ref{XTAdiabatic}) breaks down and must be
replaced, via matching at $1\ll X\ll X_{\rm break}$, by the exact
Kerr metric's adiabatic inspiral formulae \cite{finnthorne}.}

In the plunge regime, radiation reaction is unimportant; i.e., the 
orbit evolves inward with (very nearly) constant orbital angular momentum 
$\tilde L \simeq \tilde L_{\rm final}$ and energy $\tilde E \simeq
\tilde E_{\rm final}$ (which we evaluate below); i.e., the orbit is well
approximated by geodesic free fall.  In the dimensionless equation of motion
(\ref{eomdimensionless}), the free-fall approximation translates into
neglecting the last term, $T$, so 
$d^2 X / dT^2 = - X^2$, which has the analytic first integral
\begin{equation}
dX/dT = - \sqrt{{\rm constant} - {2\over3}X^3}.
\label{XTPlungePrime}
\end{equation}
For large $|X|$, the constant can be neglected and we obtain the analytic
solution
\begin{equation}
X = {- 6 \over (T_{\rm plunge} - T)^2} \quad \hbox{for plunge near the isco,}
\label{XTPlunge}
\end{equation}
which is plotted in Figs.\ \ref{figtransition1} and \ref{figtransition2}. 

Combining Eqs.\ (\ref{defxi}), 
(\ref{xitau}), 
and (\ref{XTDef}), one 
finds that
throughout the transition regime, the energy and angular-momentum 
deficits 
(i.e., the deviations of $\tilde E$ and $\tilde L$ from their isco values) 
scale as $\eta^{4/5}$.
In particular, the final deficits in the plunge stage are given by
\begin{eqnarray}
\tilde L_{\rm final}-\tilde L_{\rm isco}&=&
-(\kappa \tau_0 T_{\rm plunge}) \eta^{4/5}
\;, \nonumber\\
\tilde E_{\rm final}-\tilde E_{\rm isco}&=&
- \tilde \Omega_{\rm isco}(\kappa \tau_0 T_{\rm plunge}) \eta^{4/5}\;,
\label{final-values}
\end{eqnarray}
where, as was noted above,
\begin{equation}
T_{\rm plunge} = 3.412\;.
\label{Tplunge}
\end{equation}

\section{Gravitational Waves from Transition Regime, and their
Observability}
\label{sec:GW}

The gravitational waves emitted in the transition regime are all near 
the orbital
frequency $2\pi \Omega_{\rm isco}$ and its harmonics.  The
strongest waves are at the second harmonic (twice the orbital frequency): 
\begin{equation}
f \simeq 2 {\Omega_{\rm isco}\over2\pi} = {\tilde\Omega_{\rm isco}\over\pi M}\;.
\label{fgw}
\end{equation}
We shall compute their properties.

The transition waves last for a proper time $\Delta\tau = M \Delta\tilde\tau =
M\eta^{-1/5} \tilde \tau_o \Delta T$, 
during which the body spirals inward through 
a radial distance $\Delta r = M \Delta R = M \eta^{2/5}R_o\Delta X$, where 
$\Delta T$ covers the range $T \simeq -1$ to $\simeq 2.3$ and $\Delta X$ 
covers the range $X \simeq 1$ to $X \simeq -5$ (Fig.\ \ref{figtransition2}); 
i.e.,
\begin{equation}
\Delta T = 3.3\;, \quad \Delta X = 6\;.
\label{DeltaTDeltaX}
\end{equation}
Correspondingly, neglecting any cosmological redshift,
the duration of the transition waves as seen at Earth is
\begin{equation}
\Delta t = {M\over 
(d\tilde\tau/d\tilde t)_{\rm isco}}\eta^{-1/5}\tilde\tau_o \Delta
T\;,
\label{Deltat}
\end{equation}
and their frequency band is 
$\Delta f = (1/\pi M)(d\tilde\Omega/d\tilde r)_{\rm
isco} \Delta \tilde r$, 
which, using the above expression for $\Delta r$ and Eq.\ 
(\ref{Omega})
for $\tilde\Omega(\tilde r)$, gives
\begin{equation}
\Delta f = {3\over2\pi M}\tilde\Omega_{\rm isco}^2 \sqrt{\tilde r_{\rm isco}}
\eta^{2/5} R_o \Delta X\;.
\label{Delta f}
\end{equation}
The total number of cycles of these transition waves is
\begin{equation}
N_{\rm cyc} = f \Delta t
= {\tilde\Omega_{\rm isco} \tilde \tau_o \over \pi (d\tilde\tau/d\tilde t)_{\rm
isco}} \eta^{-1/5} \Delta T\;.
\label{Ncyc}
\end{equation}

These second-harmonic waves arriving at Earth have the form 
$h_+ = h_{+{\rm amp}}\cos(2\pi \int f dt +\varphi_+)$, 
$h_\times = h_{\times{\rm amp}}\cos(2\pi \int f dt + \varphi_\times)$,
where $\varphi_+$ and $\varphi_\times$ are constant phases.
The amplitudes $h_{+{\rm amp}}$ and $h_{\times{\rm amp}}$
depend on the source's orientation.  When one squares and adds these
amplitudes and then averages over the sky (``$\langle ... \rangle$''), 
one obtains an rms amplitude
\begin{equation}
h_{\rm amp}^{\rm rms} = \langle h_{+{\rm amp}}^2 + h_{\times{\rm
amp}}^2\rangle^{1/2}\;,
\label{hamprmsDef}
\end{equation}
which is related to the power being radiated into the second harmonic by
$\dot E_2 = 4\pi D^2 (h_{\rm amp}^{\rm rms})^2 (2\pi f)^2/(32\pi)$; cf.\ Eq.\
(35.27) of MTW \cite{mtw}.  Here $D$ is the distance to the source.
Equating this to the radiated power $\dot E_2 = (32/5) \eta^2
\tilde\Omega_{\rm isco}^{10/3} \dot {\cal E}_{\infty,2}$ 
\cite{finnthorne}, where $\dot {\cal E}_{\infty,2}$ is a relativistic
correction factor listed on the first line of Table IV of 
\cite{finnthorne}, we obtain the
following expression for the waves' rms amplitude
\begin{equation}
h_{\rm amp}^{\rm rms} = {8\over\sqrt{5}}{M\eta\over D} \tilde \Omega_{\rm
isco}^{2/3} \sqrt{\dot{\cal E}_{\infty, 2}}\;.
\label{hamprms}
\end{equation}

The signal to noise ratio $S/N$ 
that these waves produce in LISA depends on 
the orientations of LISA and the source
relative to the line of sight between them.
When one squares $S/N$ and averages over both orientations, 
then takes the square root, one obtains \cite{300yrs}
\begin{equation}
\left({S\over N}\right)_{\rm rms} = {h_{\rm amp}^{\rm rms} \over 
\sqrt{5 S_h(f)/\Delta t}}\;.
\label{SOverN}
\end{equation}
Here $5 S_h(f)$ is the spectral density of LISA's strain noise 
inverse-averaged over the sky\footnote{
i.e., $1/(5 S_h) \equiv$ average over
the sky of 1/(spectral density).  The factor $5$ in this definition is
to produce accord with the conventional notation for ground-based
interferometers, where $S_h(f)$ denotes the spectral density for waves
with optimal direction and polarization.  In the case of LISA, at frequencies
above about 0.01 Hz, the beam pattern shows sharp frequency-dependent
variations with direction due to the fact that the interferometer arms are 
acting as one-pass delay lines rather than optical cavities, and this produces
a more complicated dependence of sensitivity on angle than for ground-based
interferometers.  As a result, $S_h$ (as we have defined it)
is the spectral density for optimal direction and polarization only below
about 0.01 Hz, not above.  
} 
and $1/\Delta t$ is the band
width associated with the waves' duration $\Delta t$.

The noise spectral density $S_h(f)$ for the current straw-man design of LISA
has been computed by the LISA Mission Definition Team \cite{MDT}.
An analytic fit to this $S_h(f)$, after averaging over some small-amplitude
oscillations that occur at $f> 0.01$ Hz, is the following:
\begin{eqnarray}
S_h(f) = && \left[ (4.6\times 10^{-21})^2 + (3.5\times 10^{-26})^2\left({1 {\rm
Hz}\over f} \right)^4 \right. 
\nonumber\\
&& \quad \left. + (3.5 \times 10^{-19})^2 \left( {f\over 1 {\rm
Hz}}\right)^2 \right] {\rm Hz}^{-1}\;.
\label{ShMDT}
\end{eqnarray}

The rate for
$\mu\sim 10 M_\odot$ black holes to spiral into $M\sim 10^6 M_\odot$
black holes in galactic nuclei has been estimated by Sigurdsson and Rees
\cite{sigurdssonrees}; their ``very conservative'' result is $\sim$ one event
per year out to 1 Gpc. The inspiraling holes are likely to be in rather
eccentric, nonequatorial orbits \cite{hilsbender}, 
for which our analysis needs to be
generalized.  If, however, the orbit is circular and equatorial and 
the holes are at 1 Gpc distance, then the above formulas give the numbers
shown in Table \ref{table.tbl2}.

As shown in the table, the signal to noise for this source is of order 
unity. With some 
luck in the orientation of LISA, the orientation of the source, the
distance to the source, and/or the holes' masses,
a $S/N$ of a few might occur.  
Since the signal would already have been detected from the much stronger
adiabatic inspiral waves, this signal strength could be enough to 
begin to explore the details of the transition from inspiral to
plunge.

\section{Conclusions}
\label{sec:Conclusions}

Our analysis of the transition regime has been confined to circular, equatorial
orbits.  This is a serious constraint, since there is strong reason to expect
that most inspiraling bodies will be in orbits that are strongly noncircular
and nonequatorial \cite{hilsbender}.  
Our estimated signal-to-noise ratio, $S/N \sim 1$, 
for LISA's observations of the transition regime from a circular, 
equatorial orbit at the plausible distance $\sim 1$ Gpc 
suggests that for more realistic orbits the transition regime
{\it might} be observable.
This prospect makes it important to 
generalize our analysis to more realistic orbits.

Full analyses for equatorial, noncircular orbits and for nonequatorial, 
circular orbits can be carried out using techniques now in hand:  the Teukolsky
formalism, and computations of the orbital evolution based on
the energy and angular momentum radiated down the hole and off to infinity
(see, e.g., Ref.\ \cite{hughes} and references therein).
For nonequatorial, noncircular orbits, the analysis should
also be possible with existing techniques --- up to an unknown
radiation-reaction-induced rate of evolution of the Carter constant.  
That unknown
quantity could be left as a parameter in the analysis, to be determined
when current research on gravitational radiation reaction 
\cite{mino,quinnwald,barackori} has reached fruition.

When this paper was in near final form, we became aware of a similar 
analysis, by Buonanno and Damour \cite{buonannodamour1}, of the transition 
from inspiral to plunge.  Whereas we treat the case of infinitesimal mass
ratio $\eta \ll 1$ and finite black-hole spin $-1<a<+1$, Buananno and 
Damour treat finite $\eta$ ($0<\eta\le 1/4$)
and vanishing spins $a=0$.  Both analyses give
the same dimensionless equation of motion (\ref{eomdimensionless}) for the
transition regime.

\section*{Acknowledgments}

We thank James Anderson for helpful discussions, 
Theocharis Apostolatos and Richard O'Shaughnessy
for checking some details of our analysis, 
and Sam Finn for helpful advice and for permission to
use unpublished results from his numerical solutions of the Teukolsky
equation; see Tables \ref{Table.tbl} and \ref{table.tbl2}.  
This work was supported in part
by NASA grant NAG5-6840, and in view of its potential applications to LIGO,
also by NSF grant AST-9731698.

\newpage

\widetext

\begin{table}
\caption{Dimensionless parameters characterizing the isco and the 
transition regime of inspiral. 
The values of $\dot {\cal E}$ are from numerical solutions
of the Teukolsky equation by L.\ S.\ Finn (first line of Table II of
Ref.\ \protect\cite{finnthorne}).
}
\begin{tabular}{ddddddddd}
$a$ & $\tilde r_{\rm isco}$ & $\tilde \Omega_{\rm isco}$ & $\dot {\cal E}$ &
$\alpha$ & $\beta$ & $\kappa$ & $R_o$ & $\tau_o$ \\\tableline
-0.99& 8.972& 0.03863& 1.240& 0.0001543& 0.006626& 0.005013& 3.129& 45.50\\
-0.9& 8.717& 0.04026& 1.233& 0.0001732& 0.007070& 0.005527& 3.117& 43.04\\ 
-0.5& 7.555& 0.04935& 1.197& 0.0003070& 0.009730& 0.008966& 3.048& 32.69\\ 
0& 6.000& 0.06804& 1.143& 0.0007716& 0.01604& 0.01955& 2.925& 21.05\\
0.2& 5.329& 0.07998& 1.114& 0.001240& 0.02057& 0.02914& 2.852& 16.82\\ 
0.5& 4.233& 0.1086& 1.053& 0.003115& 0.03270& 0.06291& 2.687& 10.93\\ 
0.8& 2.907& 0.1737& 0.9144& 0.01401& 0.06446& 0.2123& 2.326& 5.539\\
0.9& 2.321& 0.2254& 0.7895& 0.03447& 0.09039& 0.4214& 2.041& 3.770\\ 
0.99& 1.454& 0.3644& 0.4148& 0.2234& 0.1289& 1.531& 1.284& 1.867\\
0.999& 1.182& 0.4379& 0.2022& 0.5127& 0.09568& 2.594& 0.8551& 1.510\\
\end{tabular}
\label{Table.tbl}
\end{table}

\begin{table}
\caption{Properties of the second-harmonic, transition-regime gravitational
waves from
a $\mu = 10 M_\odot$ black hole spiraling into a $M = 10^6 M_\odot$ black hole
(so $\eta = \mu/M = 10^{-5}$) at $r= 1$ Gpc distance.  
The values of $\dot {\cal E}_{\infty, 2}$ are from numerical solutions
of the Teukolsky equation by L.\ S.\ Finn (first line of Table IV of Ref.\  
\protect\cite{finnthorne}).}
\begin{tabular}{dddddddd}
$a$& $f$, Hz& ${\Delta f\over f}$& $\Delta t$, sec& $N_{\rm cyc}$&
$\dot {\cal E}_{\infty, 2}$& 
$h_{\rm amp}^{\rm rms}$& $\left({S\over N}\right)_{\rm rms}$, $10^{-22}$ 
\\\tableline 
-0.99& 0.002496& 0.033& 9300& 23& 1.029& 2.0& 1.2\\
-0.9& 0.002601& 0.033& 8800& 23& 1.020& 2.0& 1.2\\
-0.5& 0.003188& 0.037& 7000& 22& 0.9734& 2.3& 1.4\\
0.& 0.004396& 0.044& 4800& 21& 0.8957& 2.7& 1.6\\
0.2& 0.005167& 0.047& 4100& 21& 0.8535& 2.9& 1.6\\
0.5& 0.007016& 0.054& 2900& 21& 0.7653& 3.4& 1.6\\
0.8& 0.01123& 0.062& 1900& 22& 0.5914& 4.1& 1.3\\
0.9& 0.01457& 0.063& 1700& 24& 0.4617& 4.3& 1.1\\
0.99& 0.02354& 0.051& 1800& 43& 0.1656& 3.6& 0.72\\
0.999& 0.02829& 0.037& 3400& 96& 0.06128& 2.4& 0.58\\
\end{tabular}
\label{table.tbl2}
\end{table}

\end{document}